\documentclass[a4paper,11pt]{article}
\usepackage{amsfonts}
\usepackage{graphicx}
\usepackage[colorlinks]{hyperref}
\usepackage{amsmath}
\usepackage{amssymb}
\usepackage{latexsym}
\usepackage{color}
\usepackage{cite}
\usepackage{float}
\usepackage[textfont={footnotesize,sf},labelfont={color=blue,bf,sf},labelsep=endash]{caption}
\usepackage[position=top,labelfont={color=blue,bf,sf}]{subfig}
\usepackage{bm}
\usepackage{makeidx}
\usepackage{colortbl}
\usepackage{csquotes}
\usepackage[squaren]{SIunits}
\newcommand{\bra}{\begin{array}}
\newcommand{\era}{\end{array}}
\newcommand{\beq}{\begin{equation}}
\newcommand{\eeq}{\end{equation}}
\newcommand{\beqar}{\begin{eqnarray}}
\newcommand{\eeqar}{\end{eqnarray}}

\def\BC{\bb C}
\def\_\BC{\bbi C}

\def\Tr {{\rm Tr}}

\def\( {\left(}
   \def\) {\right)}
\def\[ {\left[}
\def\] {\right]}

\def\Tr {{\rm Tr}}
\def\dag {{\dagger}}

\newcommand{\lb}{\label}

\begin{document}
\begin{titlepage}
\setcounter{page}{1}
\renewcommand{\thefootnote}{\fnsymbol{footnote}}


\vspace{5mm}
\begin{center}

{\Large \bf {Thermodynamic Properties of q-deformed  massless Dirac fermions in  graphene with Rashba coupling  }}

\vspace{5mm}

{\bf Rachid Hou\c{c}a\footnote{\color{blue}houca.rachid@gmail.com}$^{1,2}$, El Bouâzzaoui Choubabi\footnote{\color{blue}choubabi@gmail.com}$^{2}$, Abdelhadi Belouad\footnote{\color{blue}belabdelhadi@gmail.com}$^{2}$,  Abdellatif Kamal\footnote{\color{blue}abdellatif.kamal@ensam-casa.ma}$^{2,3}$ and Mohammed El Bouziani\footnote{\color{blue}elbouziani@yahoo.com}$^{2}$}

\vspace{5mm}

{$^{1}$\em Team of Theoretical Physics and High Energy, Department of Physics, Faculty of Sciences, Ibn
Zohr University}, Agadir, Morocco,\\
{\em PO Box 8106, Agadir, Morocco}

{$^{2}$\em Team of Theoretical Physics, Laboratory L.P.M.C., Department of Physics, Faculty of Sciences, Chouaib
Doukkali University, El Jadida, Morocco},\\
{\em PO Box 20, 24000 El Jadida, Morocco}

{$^{3}$\em Department of Mechanical Engineering, National Higher School of Arts and Crafts, Hassan II
University, Casablanca, Morocco}

\vspace{3cm}

\begin{abstract}
We study the thermodynamic properties of massless Dirac fermions in graphene under
a uniform magnetic field and Rashba spin-orbit coupling with a $q-$deformed
Heisenberg algebra calculus. The thermodynamic
functions such as the Helmholtz free energy, total energy, entropy and heat capacity are obtained
by using an approach based on the zeta function and Euler-Maclaurain formula. These functions
will be numerically examined for different values of $\eta={1\over i}\ln(q)$. In particular, the heat capacity in
the  presence of deformation, all curves coincide and reach the fixed value $C=6K_B$ three times greater
compared to the case of undeformed massless Dirac fermions in graphene.
\end{abstract}

\end{center}

\vspace{4cm}



\end{titlepage}

\section{Introduction}
Graphene, which is an elemental sheet of graphite, consists of a periodic, two-dimensional arrangement of carbon atoms of monoatomic thickness with a honeycomb structure. It is the latest member of the carbon allotropic family: diamond, graphite, $C60$ fullerenes \cite{Kroto} and nanotubes \cite{Iijima}. For the first time in $2004$, a graphene sheet stable at room temperature was obtained physically by A. Geim and K. Novoselov \cite{Novoselov}. This experiment contradicted the theory that a graphene sheet was thermodynamically unstable. As this new material developed by mechanical exfoliation has remarkable and unique properties, they were awarded the Nobel Prize in Physics in $2010$. Since this discovery, graphene has been the material most studied by the scientific community for its new and unique physical properties. Indeed, it has a high superior electrical mobility \cite{Morozov,Bolotin}, an anomalous quantum Hall effect \cite{Geim}, a modulable band gap \cite{Castro} and it is a transparent conductor \cite{Nair} since, in the optical region, it absorbs only $2.3\%$ of light. It also has good flexibility \cite{Peres} and excellent mechanical strength, and its thermal conductivity is ten times higher than that of copper \cite{Ghosh}.

The more general framework of the $q-$deformation theory for a real parameter $q$ has found great success and has attracted considerable attention from physicists and mathematicians. The interesting physical application was started by the introduction of the $q-$deformed harmonic oscillator by Biedenharn \cite{Biedenharn} and Macfarlane \cite{Macfarlane} in $1989$. Quantum mechanics can be considered as a deformation (the deformation parameter is $\hbar$) of classical mechanics  and relativistic mechanics is another deformation (with $1\over c$ as the deformation parameter) of classical mechanics. In the same sense, quantum mechanics can be seen as the limit of a more general theory depending on one or more deformation parameters.

The study of the dynamic behavior of systems is a central question in physics and mathematics. These systems provide fundamental and general results which have found major applications not only in physics, but also in all other branches of science, as well as in technology; however, the harmonic oscillator is the simplest and most fundamental theoretical model of mechanical and electrical oscillatory phenomena. The definitions and main properties of independent and time dependent harmonic oscillators and damped harmonic oscillators have been studied by several authors \cite{celia,Baldiotti}.

It was lately shown that the regular standard thermodynamics of Boltzmann-Gibb's statistics are no longer suitable for studying all physical systems, including the attitude of complex systems controlled by the Tsallis non-extensive statistics \cite{Tsallis} and non-equilibrium statistics of the q-deformed superstatistics \cite{ Abe,Beck }. The concept of superstatistics was first developed by Wilk and Wlodarczyk \cite{Wilk} before Beck and Cohen \cite{ Cohen} latter reworded the theory.

Motivated by the work done on thermodynamic proprieties under magnetic field and Rashba coupling  \cite{houca} we will generalize this last work by introducing the notion of the $q$-deformed harmonic oscillator and see the influence of the parameter $q$ on the various thermodynamic quantities. For this, we consider a massless Dirac fermions in  monolayer graphene with the magnetic field  applied perpendicular to the graphene layer. Through investment of The $q$-deformed algebra of the quantum oscillator  defined by q-deformed Heisenberg algebra we express our Hamiltonian in terms of creation and annihilation operators to  obtain the solutions of the energy spectrum, this last will be used to determine the partition function which will help us to calculate and plot numerically the different thermodynamic functions in order to make conclusions. The present paper is organized as follows. In section {\color{blue}2},  we give an overview on q-deformed Heisenberg algebra which serves to determine, explicitly, the exact
eigenvalues in terms of $q$-deformed parameter. In section {\color{blue}3}, we will look for the partition function which will be the key to determine the different thermodynamic functions such as the free energy of Helmholtz, internal energy, entropy and heat capacity. section {\color{blue}4}, will be devoted to the numerical results and discussions as well as comparison with literature. We conclude our results in the final section.
\section{Theoretical model}
\subsection{q-deformed quantum theory}
The q-deformed algebra of the quantum oscillator is defined by q-deformed Heisenberg algebra in terms of creation and
annihilation operators $a^\dag$ and $a$, respectively, and number operator  $N$ by \cite{Chaichian,Ng,Sakurai}
\begin{equation}\label{qdef}
 [a,a]=[a^\dag,a^\dag]=0, \quad [a,a^\dag]_q=aa^\dag-q^{-1}a^\dag a=q^{N},\quad [N,a^\dag]=a^\dag,\quad [N,a]=-a
\end{equation}
where deformation parameter $q$ is real and the observed value of q has to satisfy the non-additivity property
\beq
[x+y]\neq [x]+[y]
\eeq
In addition, the operators obey the relations
\beq
[N]=a^\dag a,\quad [N+1]=a a^\dag
\eeq
The $q-$Fock space spanned by orthornormalized eigenstates $|n\rangle$ is constructed according to
\beq
|n\rangle={(a^\dag)^n\over\sqrt{[n]!}}|0\rangle,\quad a|0\rangle=0
\eeq
Both $q-$factorial and $q-$numbers are defined, respectively, by
\beq\lb{nn}
[n]!=[n][n-1][n-2]\cdots[1],\quad [n]={q^n-q^{-n}\over q-q^{-1}}
\eeq
For $n \in \mathbb{N}$ with $[0]!=1$.  The eigenvalues of the $q-$deformed one-dimensional harmonic oscillator are
\beq
E_n={\hbar\omega\over2}\left([n]+[n+1]\right)
\eeq
Considering the definition of basic number given in (\ref{nn}), and making $q=e^{i\eta}$, the eigenvalues become
\beq
E_n={\hbar\omega\over2}{\sin[\eta(n+{1\over2})]\over\sin[{\eta\over2}]}
\eeq
\subsection{Eigenvalue problem}
We consider a massless Dirac fermions in  monolayer graphene with the magnetic field  applied perpendicular to the graphene layer. Low energy quasiparticles in graphene with Rashba spin orbit coupling (RSOC) interaction can be well described by the Dirac-type Hamiltonian
\beq\lb{HH}
H=v_F\left(\eta\sigma_x  \pi_x+\sigma_y  \pi_y\right)+\lambda_R\left(\eta\sigma_xs_y -\sigma_ys_x \right)
\eeq
where  the conjugate momentum $\pi_x$ and $\pi_y$ can be written in symmetric gauge $\vec{A}=\frac{B}{2}(-y,x)$ as
\beq
\pi_x=p_x-\frac{eB}{2}y,\qquad \pi_y=p_y+\frac{eB}{2}x.
\eeq
Where $B$ and $v_F=10^6 m/s$ are respectively the uniform magnetic field and the Fermi velocity, the parameter $\eta = \pm1$ labels the valley degrees
of freedom, $\sigma=\left(\sigma_x,\sigma_y\right)$ are the Pauli matrices of pseudospin operator
on $A(B)$ lattice cites. The present system  presents the intrinsic spin orbit coupling
(SOC), but its value is very weak compared to the RSOC \cite{houca}, for this we have neglected it  because it will not influence on the physical properties of the studied system.

Fixing a certain intra-Landau-level quantum number, we denote by $|r_{A,B},n,\sigma\rangle={(a^\dag)^n\over \sqrt{[n]!}}|r_{A,B},n,\sigma\rangle$
a state in the \textit{nth} Landau level with spin direction $\sigma\in\{\uparrow,\downarrow\}$ , and all other eigenstates are of the form $|\Psi\rangle = ( |r_A,n,\uparrow\rangle,|r_B,n-1,\downarrow\rangle,|r_B,n,\uparrow\rangle,|r_A,n-1,\downarrow\rangle)^t$. The Hamiltonian (\ref{HH}) around a single Dirac point $(\eta = +1)$ with these considerations is given by
 \beq\lb{hhi}
H=\left(
\begin{array}{cccc}
 0 & 0 & v_F\left(\pi_x-i\pi_y \right) & 0 \\
 0 & 0 & 0 & v_F\left(\pi_x+i\pi_y \right) \\
v_F\left(\pi_x+i\pi_y \right) & 0 & 0 & -2i\lambda_R  \\
 0 & v_F\left(\pi_x-i\pi_y \right) & 2i\lambda_R  & 0 \\
\end{array}
\right).
\eeq
To diagonalize the Hamiltonian (\ref{hhi}), it is convenient to  introduce the usual bosonic operators
in terms of the conjugate momentum
\beq
a=\frac{\ell_B}{\sqrt{2}\hbar}\left(\pi_x-i\pi_y\right),\qquad a^\dag=\frac{\ell_B}{\sqrt{2}\hbar}\left(\pi_x+i\pi_y\right)
\eeq
which verify the commutation relation $[a,a^\dag]=q^{N}$,  $\ell_B = \sqrt{\frac{\hbar}{eB}}$  is the magnetic length.
Express (\ref{hhi}) in terms of $a$ and $a^\dag$ to obtain
\beq
H=\left(
 \begin{matrix}
 0 & 0 & \frac{\sqrt{2} \hbar v_F}{\ell_B} a & 0 \\
 0 & 0 & 0 & \frac{\sqrt{2} \hbar v_F}{\ell_B} a^\dag \\
 \frac{\sqrt{2} \hbar v_F }{\ell_B}a^\dag & 0 & 0 & -2i\lambda_R  \\
 0 & \frac{\sqrt{2} \hbar v_F}{\ell_B} a & 2i\lambda_R  & 0 \\
\end{matrix}
\right).
\eeq
To find the solution of the energy spectrum
we  act the Hamiltonian on the state $|\Psi\rangle$  leading to the eigenvalue equation
\beq
\left(
 \begin{matrix}
 -E & 0 & \frac{\sqrt{2} \hbar v_F}{\ell_B} a & 0 \\
 0 & -E & 0 & \frac{\sqrt{2} \hbar v_F}{\ell_B} a^\dag \\
 \frac{\sqrt{2} \hbar v_F}{\ell_B}a^\dag & 0 & -E & -2i\lambda_R  \\
 0 & \frac{\sqrt{2} \hbar v_F}{\ell_B} a & 2i\lambda_R  & -E \\
\end{matrix}
\right)\left(
         \begin{matrix}
           |r_A,n,\uparrow\rangle \\
           |r_B,n-1,\downarrow\rangle \\
           |r_B,n,\uparrow\rangle \\
           |r_A,n-1,\downarrow\rangle \\
         \end{matrix}
       \right)=\left(
                 \begin{matrix}
                   0 \\
                   0 \\
                   0 \\
                   0 \\
                 \end{matrix}
               \right)
\eeq
giving  the following system of equations
\begin{eqnarray}
 && -E|r_A,n,\uparrow\rangle+\frac{\sqrt{2} \hbar v}{\ell_B}a|r_B,n,\uparrow\rangle = 0 \\
  &&-E|r_B,n-1,\downarrow\rangle+\frac{\sqrt{2} \hbar v}{\ell_B} a^\dag|r_A,n-1,\downarrow\rangle = 0 \\
  && \frac{\sqrt{2} \hbar v}{\ell_B} a^\dag|r_A,n,\uparrow\rangle-E|r_B,n,\uparrow\rangle-2i\lambda_R|r_A,n-1,
  \downarrow\rangle = 0
  \\
  && \frac{\sqrt{2} \hbar v}{\ell_B}a|r_B,n-1,\downarrow\rangle+2i\lambda_R|r_B,n,\uparrow\rangle)-E|r_A,n-1,\downarrow\rangle =0.
\end{eqnarray}
These can be solved to obtain a second order equation for the eigenvalues
\beq
E^2\pm2\lambda_RE-+\left(\hbar\omega_D\right)^2[n]=0,\qquad n=0,1,2\cdots
\eeq
where $\omega_D=v_F\sqrt{\frac{2 eB}{\hbar }}$ is the Dirac constant. The following solutions of the last equations are the form
\begin{eqnarray}\lb{eee}
  && E_{1,n}^\pm = -\lambda_R\pm\sqrt{\left(\hbar\omega_D\right)^2[n]+\lambda_R^2} \\
  && E_{2,n}^\pm = \lambda_R\pm\sqrt{\left(\hbar\omega_D\right)^2[n]+\lambda_R^2}
\end{eqnarray}
We note that the preceding energies depend to the Rashba coupling parameter  $\lambda_R$ and the $q-$deformation parameter, now using the equation (\ref{nn}) to find
\begin{eqnarray}
  && E_{1,n}^\pm = \lambda_R\left(-1\pm\sqrt{1+\left({\hbar\omega_D\over\lambda_R}\right)^2{\sin(n\eta)\over\sin(\eta)}}\right) \\
  && E_{2,n}^\pm = \lambda_R\left(1\pm\sqrt{1+\left({\hbar\omega_D\over\lambda_R}\right)^2{\sin(n\eta)\over\sin(\eta)}}\right)
\end{eqnarray}
\begin{figure}[!ht]
  \centering
  \includegraphics[width=4.5in]{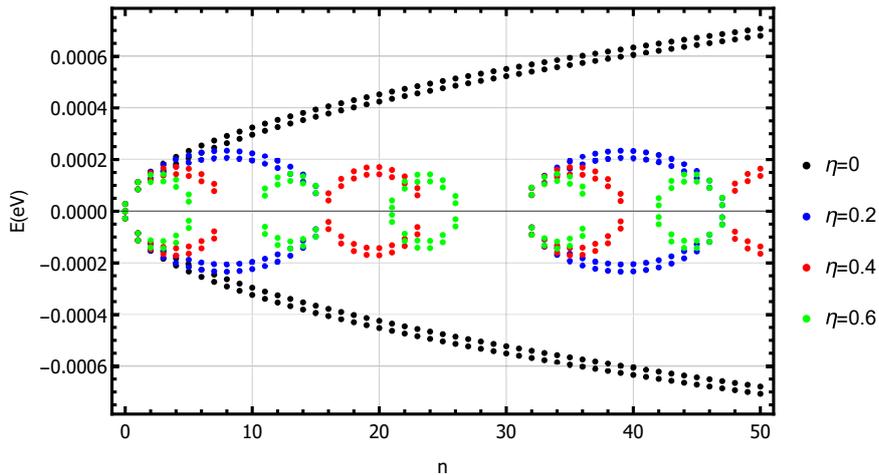}
  \caption{(Color online) Eigenvalue $E$ versus $n$ for different values of $q-$deformed parameters, $\eta=0, 0.2, 0.4, 0.6$, respectively for the values of the magnetic field and Rashba coupling parameter $B\sim10^{-3}$ and
    $\lambda_R=0.014 meV$ \cite{houca}.  }\label{yy}
\end{figure}
In Figure (\ref{yy}), we present the eigenvalues of the $q-$deformed massless Dirac fermions in graphene, thus, the energy
levels shows that when there is no deformation the energy  is quantified and has a parabolic form and symmetrical compared to the quantization axis $n$, but for a given deformation we notice that the parabolic form tends towards an ellipse, on the other hand there is an appearance of a second quantification via the periodicity of ellipses.

Now if we consider very small deformation and neglect all terms proportional to $\eta^4$, we have
\begin{eqnarray}\lb{vv}
  && E_{1,n}^\pm =\epsilon^\pm_{1,n}\pm\Delta\epsilon(n) \\ \nonumber
  && E_{2,n}^\pm =\epsilon^\pm_{2,n}\pm\Delta\epsilon(n)
\end{eqnarray}
where $\varrho$, $\epsilon^\pm_{1,n}$, $\epsilon^\pm_{2,n}$ and $\Delta\epsilon(n)$ are defined by

\begin{eqnarray}
  && \epsilon^\pm_{1,n} = \lambda_R\left[-1\pm\sqrt{1+\varrho^2n}\right] \\
  && \epsilon^\pm_{2,n} = \lambda_R\left[1\pm\sqrt{1+\varrho^2n}\right]\\
  && \Delta\epsilon_n =-\frac{\lambda_R\varrho^2}{12}\frac{n^3}{\sqrt{1+\varrho^2n}}\eta^2 \\
  && \varrho ={\hbar\omega_D\over\lambda_R}
\end{eqnarray}
The term $\Delta\epsilon_n$ is the correction on the energy when the deformation exists, without deformation i.e. $ q\rightarrow 1 $ ($ \eta\rightarrow 0 $)  the last energies are reduced to the expressions already found in \cite{houca}.

\section{Thermodynamic quantities}
We will study the thermodynamic properties of massless Dirac fermions in graphene with Rashba coupling in contact
with a thermal reservoir at finite temperature. For  simplicity, we assume that only fermions with positive energy
$(E > 0)$ are regarded to constitute the thermodynamic ensemble \cite{houca}. We start by evaluating
\beq\lb{ZZe}
\mathbb{Z}=\Tr e^{-\beta H}=
\sum_{n=0}^{+\infty}\left(e^{-\beta E_{1,n}^+}+e^{-\beta E_{2,n}^+}\right)
\eeq
where $\beta=\frac{1}{k_BT}$, $k_B$ is the Boltzmann constant and $T$ is the equilibrium temperature.
Using (\ref{vv}-\ref{ZZe}), we show that $\mathbb{Z}$ takes the form
\begin{eqnarray}
\mathbb{Z}&=&\sum_{n=0}^{+\infty}e^{-\beta\left(\epsilon^+_{1,n}+ \Delta\epsilon_n\right)}+e^{-\beta\left(\epsilon^+_{2,n}+ \Delta\epsilon_n\right)} \\ \nonumber
&=& \sum_{n=0}^{+\infty}e^{\beta\Delta\epsilon_n}e^{-\beta\left(\epsilon^+_{1,n}+\epsilon^+_{2,n}\right)}\\ \nonumber
\end{eqnarray}
noting here that the term $e^{-\beta\Delta\epsilon_n}$ is very small than $1$ then the development limit around $0$ give
\begin{eqnarray}
\mathbb{Z}&\simeq& \sum_{n=0}^{+\infty}\left(1+\beta\Delta\epsilon_n\right)e^{-\beta\left(\epsilon^+_{1,n}+\epsilon^+_{2,n}\right)}\\ \nonumber
\mathbb{Z}&\simeq&\mathbb{Z}_0+\mathbb{Z}_1
\end{eqnarray}
with the partitions functions of no deformed system $\mathbb{Z}_0$ and the correction partition function $\mathbb{Z}_1$ have as expressions
\begin{eqnarray}
&& \mathbb{Z}_0=\sum_{n=0}^{+\infty}e^{-\beta\left(\epsilon^+_{1,n}+\epsilon^+_{2,n}\right)} \\ \nonumber
&& \mathbb{Z}_1=\sum_{n=0}^{+\infty}\beta\Delta\epsilon_ne^{-\beta\left(\epsilon^+_{1,n}+\epsilon^+_{2,n}\right)}
\end{eqnarray}
the partition function $\mathbb{Z}_0$ for no deformed system is already calculated in \cite{houca} and it has the following expression
\beq
\mathbb{Z}_0=\left[{2\over \varrho^2}\left(\tau^2-1\right)+1\right] \cosh {1\over\tau }.
\eeq
Where $\tau={k_BT\over\lambda_R}$ is the reduced temperature. Indeed, the second term can be evaluated by using the Euler-Maclaurin formula; starting by the equation
\beq
\mathbb{Z}_1=\frac{\varrho^2\eta^2}{6\tau}\cosh{1\over\tau }\sum_{n=0}^{+\infty}\frac{n^3}{\sqrt{1+\varrho^2n}}
e^{-{1\over\tau }\sqrt{1+\varrho^2n}}
\eeq
 to solve the last sum it's convenient to approximate integrals by finite sums, or conversely to evaluate finite sums and infinite series using integrals, indeed we put
 \beq
 f(x)=\frac{x^3}{\sqrt{1+\varrho^2x}}e^{-{1\over\tau }\sqrt{1+\varrho^2x}}
 \eeq
using the Euler–Maclaurin formula
\beq
\sum_{x=0}^{+\infty}f(x)={1\over2}f(0)+\int_{0}^{+\infty}f(x)-\sum_{p=1}^{+\infty}{B_{2p}\over2p!}f^{(2p-1)}(0)
\eeq
 $B_{2p}$ are the Bernoulli numbers, and $f^{(2p-1)}$ is the derivative of order $(2p-1)$. Up to $p=1$, the values $f(0)$ and $f^{(1)}(0)$ are nulls, then with the straightforward calculation  the final form of $\mathbb{Z}_1$ have the form
\beq
\mathbb{Z}_1={16\eta^2\over\varrho^6}\left(15\tau^6+15\tau^5+6\tau^4+
\tau^3\right)e^{-{1\over\tau }}\cosh{1\over\tau}
\eeq
Finally, the compact final form of the q-deformed partition function of the system
\beq\lb{zzz}
\mathbb{Z}=\left({2\over \varrho^2}\left(\tau^2-1\right)+1 +{16\eta^2\over\varrho^6}\left(15\tau^6+15\tau^5+6\tau^4+
\tau^3\right)e^{-{1\over\tau }}\right)\cosh{1\over\tau}
\eeq
Since we have inferred the partition function of our framework, we would now be able to determine all related thermodynamic quantities.
The determination of all thermal properties,
such as the  Helmholtz free energy $F$, internal energy $U$, heat capacity $C$ and entropy $S$, can be obtained through the expression of the partition
function $\mathbb{Z}$ by using the following relations \cite{houca}:
\begin{eqnarray}\lb{ter}
  F &=& -\lambda_R\tau\ln \mathbb{Z} \\ \nonumber
  U &=& \lambda_R\tau^2\frac{\partial\ln \mathbb{Z}}{\partial \tau} \\ \nonumber
  \frac{S}{k_B} &=&-{1\over\lambda_R}\frac{\partial F}{\partial \tau}\\ \nonumber
  \frac{C}{k_B}&=&{1\over\lambda_R}\frac{\partial U}{\partial \tau}.
\end{eqnarray}
Then, we will numerically investigate the above thermodynamic functions to underline the conduct of our framework. This will be finished by giving a few plots under reasonable conditions and making various discussions.
\section{Numerical Results and discussions}
To make a reference to reality of graphene, we restrict our study to the low-energy regime, which may be reached by fixing an appropriates values of the Rashba coupling parameter $\lambda_R$ and the external magnetic field $B$.  Indeed, for $B\simeq10^{-3}T$ and $\lambda_R=0.014 meV$.  The thermodynamic functions versus the reduced  temperature $\tau$ for the fixed values of $\eta = 0 , 0.2 , 0.4 , 0.6, 0.8, 0.9$.

\begin{figure}[!ht]
  \centering
  \includegraphics[width=7in]{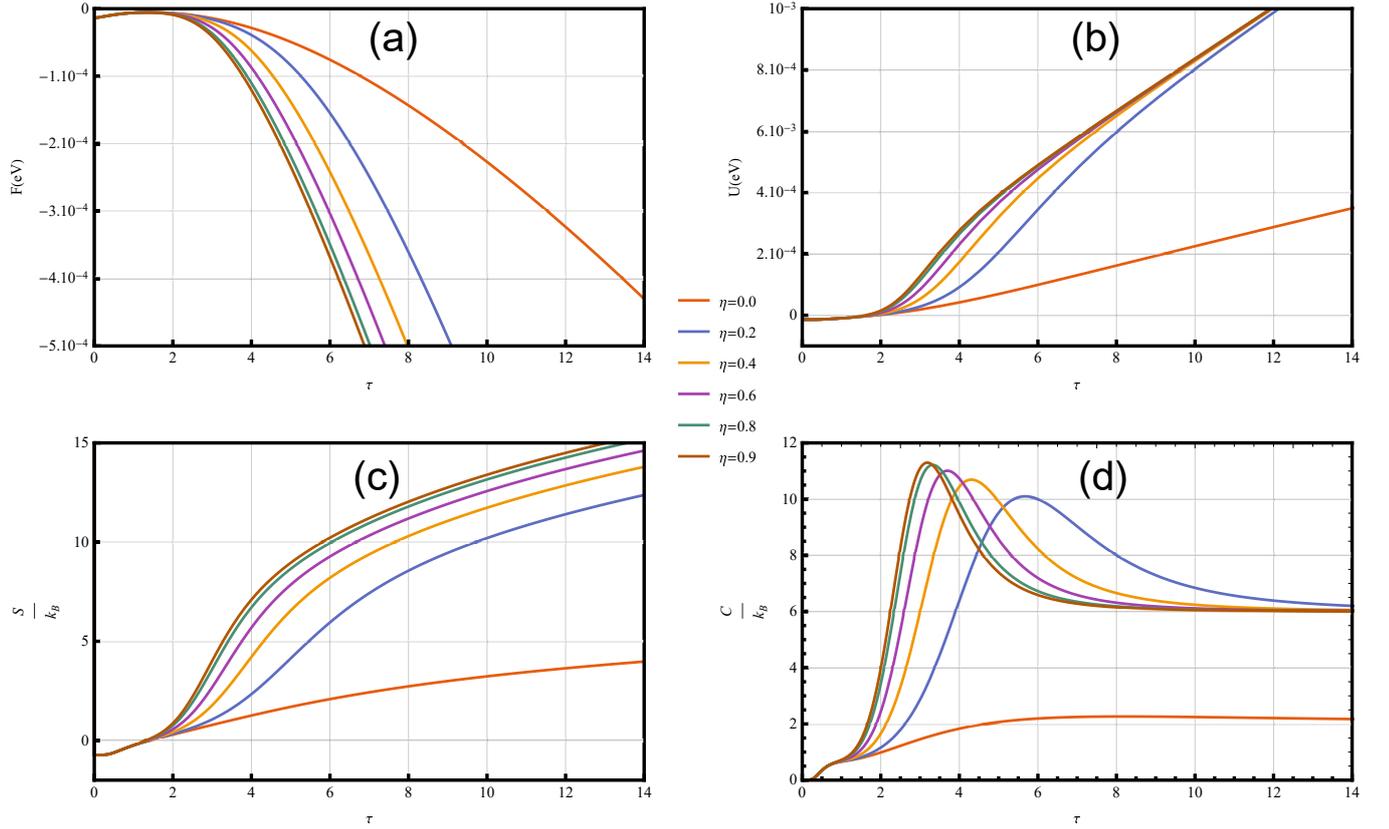}
  \caption{(Color online) Thermodynamic functions of $q-$deformed Dirac fermions in graphene with
  Rashba coupling versus the reduced temperature $\tau$  for different values of the $q-$deformed
  parameter $\eta = 0.0, 0.2, 0.4, 0.6, 0.8, 0.9$, respectively for the values of the magnetic field and Rashba coupling parameter $B\sim10^{-3}T$ and
    $\lambda_R=0.014 meV$ \cite{houca}.}\label{ff}
\end{figure}

It is clearly seen that the common remark between the four curves is the $\eta$-deformed parameter does not influence on the thermodynamic properties of the system in the low temperature regime. In figure (\ref{ff}.a) The free energy $F$ decreases gradually with increasing of temperature at a given $\eta$-deformed parameter and decreases with $\eta$ at a given temperature. In figure (\ref{ff}.b) we observe that at high temperature our system follows Joule's first law in both cases, with and without deformation, thus  in the case where the $\eta-$deformed parameter is not zero the internal energy in this regime is asymptotic to $U=6\lambda_R\tau$, but when  $\eta=0$ the internal energy become asymptotic to $U=2\lambda_R\tau$, then we observe that for two cases the internal energy in high temperature regime depends only on the reduced temperature $\tau$, then  we conclude for two cases the kinetic energy of translation of molecules is the unique form of energy of $N$ atoms contained in a volume $V$ of the system.  In figure (\ref{ff}.c) there are two remarks to report in low temperature in particular for $0<{S\over k_B}<1.7$ the entropy is negative which can be explained by the less disorder of the system \cite{houca}, in the case where ${S\over k_B}>1.7$ the entropy increases when $\eta$-parameter increases. For the tree curves at the top (a,b,c) we deduce that the parameter $\eta$-parameter  plays the same role of the doping of the graphene, however when $\eta$ increase the thermodynamic properties such as entropy, internal energy increases with $\eta$-parameter in the same way when we dope pure graphene with boron atoms $B$ or nitrogen $N$ and vice versa \cite{Mann}, and for the free energy of Helmholtz, it decrease when $\eta$ decrease  similarly when  the concentration of the doped atoms in graphene decrease. What is remarkable is in figure (\ref{ff}.d) we observe that without q-deformation our system at high temperature obeys to the Dulong-Petit law, but when the q-deformation is introduced, the heat capacity passes through a maximum in low temperature regime, that is, the point where the temperature changes very little as energy is supplied to the system, most of the energy is used to excite the carbon atoms of the ground state in the excited state, rather than increasing the kinetic energy of the system, that on the one hand, on the other hand at high temperature the heat capacity coincide and reach the fixed value $C=6K_B$ three times greater compared to the case of no deformed massless Dirac fermions in graphene which can be explained by the increasing of degree of freedom of the system due to the introduction of the $\eta$-deformed  parameter.
\section{Conclusion}
In this paper, after a brief insight on the notion of the q-deformed harmonic oscillator, we have studied the thermodynamic properties of Dirac fermions in graphene in this deformation formalism, we have found the eigenvalues of the considered system via q-deformed annihilations and creations operators. It was shown that the eigenvalues of our system are more general than in the case where there is no deformation, and especially we tested them in the limiting case $\eta=0$ where the ordinary results were well recovered. The eigenvalues are used together with a method based on the zeta function and Euler-Maclaurain formula to determine the partition function according to the q-deformed parameter. Therefore the thermodynamic functions, such as the Helmholtz free energy, total energy, entropy and heat capacity, were obtained in terms of the q-deformed parameter.

Subsequently, some cases were studied related to the $q$-deformed parameter. Indeed, we numerically analyzed the plotted curves which allowed us to make important remarks on the influence of deformation on the thermodynamic properties of our system. We also found a similarity between the doping concentration and the q-deformed parameter for the graphene system \cite{Mann}. Finally, it was shown that the Dulong-Petit law is no longer verified when the q-deformed harmonic oscillator notion is introduced where the heat capacity at high temperature tends to a constant value $C=6k_B$ three times greater in comparison with the Dirac fermions in graphene \cite{houca}.

\end{document}